\documentclass[twocolumn,english,aps,pra,superscriptaddress,showpacs]{revtex4}
\usepackage[T1]{fontenc}
\usepackage[latin9]{inputenc}
\usepackage{amstext}
\usepackage{amssymb}
\usepackage{graphicx}
\usepackage{esint}

\makeatletter
\@ifundefined{textcolor}{}
{%
 \definecolor{BLACK}{gray}{0}
 \definecolor{WHITE}{gray}{1}
 \definecolor{RED}{rgb}{1,0,0}
 \definecolor{GREEN}{rgb}{0,1,0}
 \definecolor{BLUE}{rgb}{0,0,1}
 \definecolor{CYAN}{cmyk}{1,0,0,0}
 \definecolor{MAGENTA}{cmyk}{0,1,0,0}
 \definecolor{YELLOW}{cmyk}{0,0,1,0}
}


\makeatother

\usepackage{babel}
\begin{document}

\title{Majorization approach to entropic uncertainty relations for coarse-grained
observables}

\author{\L ukasz Rudnicki}

\email{rudnicki@cft.edu.pl}

\affiliation{Institute for Physics, University of Freiburg, Rheinstra{\ss}e 10, D-79104
Freiburg, Germany}

\affiliation{Center for Theoretical Physics, Polish Academy of Sciences, Aleja
Lotnik\'ow 32/46, PL-02-668 Warsaw, Poland}
\begin{abstract}
We improve the entropic uncertainty relations for position and momentum
coarse-grained measurements. We derive the continuous, coarse-grained
counterparts of the discrete uncertainty relations based on the concept
of majorization. The obtained entropic inequalities involve two R{\'e}nyi
entropies of the same order, and thus go beyond the standard scenario
with conjugated parameters. In a special case describing the sum of
two Shannon entropies the majorization-based bounds significantly
outperform the currently known results in the regime of larger coarse
graining, and might thus be useful for entanglement detection in continuous
variables. 
\end{abstract}
\maketitle
\section{Introduction}

The optimal entropic uncertainty relation for a couple of conjugate
continuous variables (position and momentum) is known for almost 40
years \cite{BBM}. One decade later, entropic formulation of the uncertainty
principle has as well been developed in the discrete settings \cite{Deutsch,MU}.
Even though, the topic of entropic uncertainty relations (EURs) has
a long history (for a detailed review see \cite{Wehner,IBBLR}), one
can observe a recent increase of interest within the quantum information
community leading to several improvements \cite{deVicente,deVicenteComm,Partovi,my,oni,Coles,RPZ,Korzekwa1,Bosyk1,Bosyk2,Bosyk3,Kaniewski}
or even a deep asymptotic analysis of different bounds \cite{Banach}.
This is quite understandable, because the entropic uncertainty relations
have various applications, for example in entanglement detection \cite{Lewenstein,Partovi2,Rastegin1,ConEnt1,ConEnt2},
security of quantum protocols \cite{crypto,crypto2}, quantum memory
\cite{Berta,memory2} or as an ingredient of Einstein\textendash Podolsky\textendash Rosen
steering criteria \cite{Steering1,EPR}. Moreover, the recent discussion
\cite{HeisWerner} about the original Heisenberg idea of uncertainty,
led to the entropic counterparts of the noise-disturbance uncertainty
relation \cite{HeisEntropic,Rastegin2} (also obtained with quantum
memory \cite{ColFur}). 

My favorite example of entropic description of uncertainty \cite{partoviOld,IBB1,IBB2}
is situated in between the continuous and the discrete scenario. Continuous
position and momentum variables, while studied with the help of coarse-grained
measurements lead to discrete probability distributions. This particular
formulation of the uncertainty principle has been long ago recognized
\cite{Lenard,Lahti,Lahti2} to faithfully capture the spirit of position-momentum
duality. It also carries a deep physical insight, since the coarse-grained
version of the Heisenberg uncertainty relation is non-trivial for
any coarse-graining (given in terms of two widths $\Delta$ and $\delta$
in positions and momenta respectively) provided that the both widths
are finite \cite{OptCon}. On the practical level, coarse-grained
entropic relations are experimentally useful for entanglement \cite{ConEnt2,Megan}
and steering detection \cite{Steering1} in continuous variable schemes.
The aim of this paper is thus to strengthen the theoretical and experimental
tools based on the coarse-grained EURs by taking an advantage of the
recent improvements of discrete entropic inequalities, in particular,
the one based on majorization \cite{RPZ}.

Let me start with a brief description of the entropic uncertainty
landscape, with a special emphasis on the majorization approach developed
recently. The standard position-momentum scenario deals with the sum
of the continuous Shannon (or in general R{\'e}nyi) entropies $-\int dz\rho\left(z\right)\ln\rho\left(z\right)$
calculated for both densities $\rho\left(x\right)=|\psi\left(x\right)|^{2}$
and $\tilde{\rho}\left(p\right)=|\tilde{\psi}\left(p\right)|^{2}$
describing positions and momenta respectively. The position and momentum
wave functions are mutually related by the Fourier transformation.
The discrete EURs rely on the notion of the R{\'e}nyi entropy of order
$\alpha$
\begin{equation}
H_{\alpha}\left[P\right]=\frac{1}{1-\alpha}\ln\sum_{i}P_{i}^{\alpha},
\end{equation}
and the sum-inequalities of the general form
\begin{equation}
H_{\alpha}\left[P\left(A;\varrho\right)\right]+H_{\beta}\left[P\left(B;\varrho\right)\right]\geq B_{\alpha\beta}\left(A,B\right)\label{EURgen}
\end{equation}
valid for any density matrix $\varrho$, and two non-degenerate observables
$A$ and $B$. If by $\bigl|a_{i}\bigr\rangle$ and $\bigl|b_{j}\bigr\rangle$
we denote the eigenstates of the two observables in question, the
associated probability distributions entering (\ref{EURgen}) are:
\begin{equation}
P_{i}\left(A;\varrho\right)=\bigl\langle a_{i}\bigr|\varrho\bigl|a_{i}\bigr\rangle\;\;\textrm{and }\; P_{j}\left(B;\varrho\right)=\bigl\langle b_{j}\bigr|\varrho\bigl|b_{j}\bigr\rangle.\label{distributions}
\end{equation}
The lower bound $B_{\alpha\beta}$ does not depend on $\varrho$,
but only on the unitary matrix $U_{ij}=\bigl\langle a_{i}\bigr|b_{j}\bigr\rangle$.
For instance, the most recognized result by Maassen and Uffink \cite{MU}
gives the bound $-2\ln\max_{i,j}|U_{ij}|$, valid whenever 
\begin{equation}
\frac{1}{\alpha}+\frac{1}{\beta}=2.\label{conjugate}
\end{equation}
The couple $(\alpha,\beta)$ constrained as in Eq. (\ref{conjugate})
is often referred to as the conjugate parameters. 

\subsection{Majorization entropic uncertainty relations} \label{Secmajrev}

In the majorization approach one looks for the probability vectors
$Q\left(A,B\right)$ and $W\left(A,B\right)$ which majorize the tensor
product \cite{my,oni} and the \emph{direct sum} \cite{RPZ} of the
involved distributions (\ref{distributions}):
\begin{equation}
P\left(A;\varrho\right)\otimes P\left(B;\varrho\right)\prec Q\left(A,B\right),\label{maj1}
\end{equation}
\begin{equation}
P\left(A;\varrho\right)\oplus P\left(B;\varrho\right)\prec\{1\}\oplus W\left(A,B\right).\label{maj2}
\end{equation}
The majorization relation $x\prec y$ between any two $D$-dimensional
probability vectors implies that for all $n\leq D$ we have $\sum_{k=1}^{n}x_{k}^{\downarrow}\leq\sum_{k=1}^{n}y_{k}^{\downarrow}$,
with a necessary equality when $n=D$. In agreement with the usual
notation, the symbol $\downarrow$ denotes the decreasing order, what
means that $\bigl(x^{\downarrow}\bigr){}_{k}\geq\bigl(x^{\downarrow}\bigr){}_{l}$,
for all $k\leq l$.  In the case when the vectors compared in (\ref{maj1}) and (\ref{maj2}) are of different size, the shorter vector shall be completed by a proper number of coordinates equal to $0$.  The tensor product $x\otimes y$ (also called
the Kronecker product) is a $D^{2}$-dimensional probability vector
with the coefficients equal to 
\begin{equation}
x_{1}y_{1},x_{1}y_{2},\ldots,x_{1}y_{D},\ldots,x_{D}y_{1},x_{D}y_{2},\ldots,x_{D}y_{D},
\end{equation}
while the direct sum $x\oplus y$ is a $2D$-dimensional probability
vector given by
\begin{equation}
x_{1},x_{2},\ldots,x_{D},y_{1},y_{2},\ldots,y_{D}.
\end{equation}

One of the most important properties of the R{\'e}nyi entropy of any order
$\alpha$ is its additivity 
\begin{equation}
H_{\alpha}\left[x\right]+H_{\alpha}\left[y\right]=H_{\alpha}\left[x\otimes y\right].\label{add1}
\end{equation}
Moreover, in the special case of the Shannon entropies ($H_{1}\left[\cdot\right]\equiv H\left[\cdot\right]$)
one easily finds that
\begin{equation}
H\left[x\right]+H\left[y\right]=H\left[x\oplus y\right].\label{add2}
\end{equation}
Since the presumed majorization relations (\ref{maj1}, \ref{maj2})
are valid for every $\varrho$, the Schur-concavity of the R{\'e}nyi (Shannon)
entropy together with (\ref{add1}, \ref{add2}) immediately lead
to the corresponding bounds $B_{\alpha\alpha}=H_{\alpha}\left[Q\right]$
\cite{my,oni} and $B_{11}=H\left[W\right]$ \cite{RPZ}. Due to the
subadditivity property of the function $\ln(1+z)$ the validity of
the latter bound can be extended \cite{RPZ} to the range $\alpha\leq1$
(in that case the function $\sum_{i}z_{i}^{\alpha}$ is as well Schur-concave),
i.e. $B_{\alpha\alpha}^{\alpha\leq1}=H_{\alpha}\left[W\right]$. On
the other hand, when $\alpha>1$, this bound can be appropriately
modified to the weaker form \cite{RPZ}
\begin{equation}
B_{\alpha\alpha}^{\alpha>1}\left[W\right]=\frac{2}{1-\alpha}\left[\ln\left(1+\sum_{i}W_{i}\right)-\ln2\right].\label{modified}
\end{equation}

The whole families of the vectors $Q\left(A,B\right)$ and $W\left(A,B\right)$
fulfilling (\ref{maj1}) and (\ref{maj2}) have been explicitly constructed
in \cite{my,oni} and \cite{RPZ} respectively. The aim of the present
paper is to obtain the counterpart of the majorizing vector $W\left(A,B\right)$
applicable to the position-momentum coarse-grained scenario described
in detail in the forthcoming Section \ref{subsec1B}. In Section \ref{SecionMAJ}
we derive this vector using the sole idea of majorization, so that
we shall omit here a detailed prescription established in \cite{RPZ}.
We restrict the further discussion to the direct-sum approach, since
for $\alpha\leq1$ (this case covers the sum of two Shannon entropies),
the\emph{ direct-sum entropic uncertainty relation} is always stronger
than the corresponding tensor-product EUR \cite{RPZ}. 

\subsection{Entropic uncertainty relations for coarse-grained observables}
\label{subsec1B}

The last set of ingredients we shall introduce, contains the coarse-grained
probabilities together with their EURs. Due to coarse-graining, the
continuous densities $\rho\left(x\right)$ and $\tilde{\rho}\left(p\right)$
become the discrete probabilities: 
\begin{equation}
q_{k}^{\Delta}=\int_{k^{-}\Delta}^{k^{+}\Delta}dx\,\rho\left(x\right),\quad p_{l}^{\delta}=\int_{l^{-}\delta}^{l^{+}\delta}dp\,\tilde{\rho}\left(p\right),\label{rs}
\end{equation}
with $k^{\pm}=k\pm1/2$, $l^{\pm}=l\pm1/2$ and $k,l\in\mathbb{Z}$.
The sum of the R{\'e}nyi entropies $H_{\alpha}\bigl[q^{\Delta}\bigr]$
and $H_{\beta}\bigl[p^{\delta}\bigr]$ calculated for the probabilities
(\ref{rs}) is lower-bounded by \cite{OptCon} 
\begin{equation}
B_{\alpha\beta}\left(\Delta,\delta\right)=\max\left[\mathcal{B}_{\alpha}\left(\Delta\delta/\hbar\right);\mathcal{R}\left(\Delta\delta/\hbar\right)\right],\label{Boundy}
\end{equation}
where \cite{IBB2} 
\begin{equation}
\mathcal{B}_{\alpha}\left(\gamma\right)=-\frac{1}{2}\left(\frac{\ln\alpha}{1-\alpha}+\frac{\ln\beta}{1-\beta}\right)-\ln\left(\gamma/\pi\right),\label{IBB}
\end{equation}
and \cite{OptCon}
\begin{equation}
\mathcal{R}\left(\gamma\right)=-\ln\left(\gamma/2\pi\right)-\!2\ln R_{00}\left(\gamma/4,1\right)\!\geq0.\label{LR}
\end{equation}
Once more the above results are valid only for conjugate parameters
(\ref{conjugate}), so that we label the bound (\ref{IBB}) only by
the index $\alpha$. The function $R_{00}\left(\xi,\eta\right)$ is
the ``00'' radial prolate spheroidal wave function of the first
kind \cite{abr}. When $\gamma\ll1$, the spheroidal term in (\ref{LR})
becomes negligible and we have 
\begin{equation}
\mathcal{R\left(\gamma\right)}\approx\mathcal{B}_{1}\left(\gamma\right)+\ln2-1,
\end{equation}
so that the bound (\ref{IBB}) dominates in this regime. In the opposite
case, when $\gamma>e\pi\approx8.54$ the bound (\ref{IBB}) is negative,
so starting from some smaller ($\alpha$-dependent) value of $\gamma$
the second bound $\mathcal{R\bigl(\gamma\bigr)}$ becomes significant.

\section{Direct-sum majorization for coarse-grained observables}\label{SecionMAJ}

After the short but comprehensive introduction, we are in position
to formulate the main result of this paper. Assume that a sum of any
$M$ position probabilities $q^{\Delta}$ and any $N$ momentum probabilities
$p^{\delta}$ is bounded by $1+G_{MN}\left(\gamma\right)$, that is
($\gamma=\Delta\delta/\hbar$) 
\begin{equation}
q_{k_{1}}^{\Delta}+\ldots+q_{k_{M}}^{\Delta}+p_{l_{1}}^{\delta}+\ldots+p_{l_{N}}^{\delta}\leq1+G_{MN}\left(\gamma\right),\label{G}
\end{equation}
for some indices $k_{1}\neq k_{2}\neq\ldots\neq k_{M}$ and $l_{1}\neq l_{2}\neq\ldots\neq l_{N}$.
We implicitly assume here that $G_{MN}\left(\gamma\right)$ does not
depend on the specific choice of the probabilities in the sum (it
bounds any choice), and that $G_{MN}\left(\gamma\right)\leq1$ since
the left hand side of (\ref{G}) cannot exceed $2$. Denote further
by
\begin{equation}
F_{J}\left(\gamma\right)=\max_{0\leq M\leq J}G_{M,J-M}\left(\gamma\right).\label{F}
\end{equation}
Assume now that $F_{J}\left(\gamma\right)$, $J=1,2,\ldots\infty$
is an increasing sequence 
\begin{equation}
F_{J+1}\left(\gamma\right)\geq F_{J}\left(\gamma\right).\label{inc}
\end{equation}
If that happens, the construction of the vector $W\left(\gamma\right)$
applicable to the direct-sum majorization relation, i.e. such that
$q^{\Delta}\oplus p^{\delta}\prec\{1\}\oplus W\left(\gamma\right)$
can be patterned after \cite{RPZ}: 
\begin{equation}
W_{i}\left(\gamma\right)=F_{i+1}\left(\gamma\right)-F_{i}\left(\gamma\right),\label{Wi}
\end{equation}
for $i=1,2,\ldots\infty$. Due to (\ref{inc}) the coefficients $W_{i}\left(\gamma\right)$
are all non-negative, so that they form a probability vector. Note
that $F_{1}\left(\gamma\right)\equiv0$, since one picks up only a
single probability ($M=1$, $N=0$ or $M=0$, $N=1$), and that $F_{\infty}\left(\gamma\right)\equiv1$,
because whenever the quantum state is localized (in position or momentum)
in a single bin, the left hand side of (\ref{G}) is equal to $2$.
This is in accordance with an expectation that $W\left(\gamma\right)$
is the probability vector. 

One can check by a direct inspection that
\begin{equation}
1+\sum_{i=1}^{J-1}W_{i}^{\downarrow}\bigl(\gamma\bigr)\geq1+\sum_{i=1}^{J-1}W_{i}\bigl(\gamma\bigr)=1+F_{J}\left(\gamma\right),
\end{equation}
what together with (\ref{G}) and (\ref{F}) is the essence of majorization.
As in the case of discrete majorization \cite{my,RPZ}, there is a
whole family (labeled by $n=2,\ldots\infty$) of majorizing vectors
$W^{(n)}\left(\gamma\right)$ given by the prescription $W_{i}^{(n)}\equiv W_{i}$
for $i<n$, $W_{n}^{(n)}=1-F_{n}$, and $W_{i}^{(n)}\equiv0$ when
$i>n$. In that notation, the basic vector (\ref{Wi}) is equivalent
to $W^{(\infty)}\left(\gamma\right)$, and the following majorization
chain does hold
\begin{equation}
W^{(2)}\!\succ W^{(3)}\!\succ\ldots\!\succ W^{(n)}\!\succ W^{(n+1)}\!\succ\ldots\!\succ W^{(\infty)}\equiv W.
\end{equation}

The remaining task is to find the candidates for the coefficients
$F_{J}\left(\gamma\right)$. To this end we shall define two sets:
\begin{equation}
X\left(\Delta\right)=\bigcup_{a=1}^{M}\left[k_{a}^{-}\Delta,k_{a}^{+}\Delta\right],\qquad Y\left(\delta\right)=\bigcup_{b=1}^{N}\left[l_{b}^{-}\Delta,l_{b}^{+}\Delta\right],
\end{equation}
which are simply the unions of intervals associated with the probabilities
present in (\ref{G}). The measures of these sets are equal to $M\Delta$
and $N\delta$ respectively. Eq. (\ref{G}) rewritten in terms of
the above sets simplifies to the form
\begin{equation}
\int_{X\left(\Delta\right)}dx\,\rho\left(x\right)+\int_{Y\left(\delta\right)}dp\,\tilde{\rho}\left(p\right)\leq1+G_{MN}\left(\gamma\right).
\end{equation}
Following Lenard \cite{Lenard}, we shall further introduce two projectors
$\hat{\mathcal{Q}}$ and $\hat{\mathcal{P}}$, such that for any function
$f\left(x\right)$, the function $\bigl(\hat{\mathcal{Q}}f\bigr)\left(x\right)$
has its support equal to $X\left(\Delta\right)$ and the Fourier transform
of the function $\bigl(\hat{\mathcal{P}}f\bigr)\left(x\right)$ is
supported in $Y\left(\delta\right)$. If both $X\left(\Delta\right)$
and $Y\left(\delta\right)$ are intervals, then according to Theorem
4 from \cite{Lahti2} (this theorem in fact formalizes the content
of Eq. 17 from \cite{Lahti}) the formal candidate for $G_{MN}\left(\gamma\right)$
is the square root of the largest eigenvalue $\lambda_{0}$ of the
compact, positive operator $\hat{\mathcal{Q}}\hat{\mathcal{P}}\hat{\mathcal{Q}}$.
Due to Proposition 11 (including the discussion around it) from \cite{Lenard},
the above statement remains valid for any sets $X\left(\Delta\right)$
and $Y\left(\delta\right)$. As concluded by Lenard, this is a generalization
of the seminal results by Landau and Pollak \cite{LandauPollak},
who for the first time quantified uncertainty using spheroidal functions.
It however happens \cite{Price}, that $\lambda_{0}$ has the largest
value exactly in the interval case, so that it can always be upper
bounded by the eigenvalue found by Landau and Pollak:
\begin{equation}
\lambda_{0}\leq\frac{\xi}{2\pi\hbar}\left[R_{00}\left(\xi/4\hbar,1\right)\right]^{2},\label{17}
\end{equation}
with $\xi$ being the product of the measures of the two sets in question,
that is $\xi=\bigl(M\Delta\bigr)\bigl(N\delta\bigr)$. Since the right
hand side of (\ref{17}) is an increasing function of $\xi$, we can
easily find the maximum in (\ref{F}). The maximal value of $MN$
with fixed $M+N$ is given by possibly equal contributions of the
both numbers. Since $M$ and $N$ are integers we finally get 
\begin{equation}
F_{J}\left(\gamma\right)=\sqrt{\frac{\gamma\left\lceil J/2\right\rceil \left\lfloor J/2\right\rfloor }{2\pi}}R_{00}\left(\frac{\gamma\left\lceil J/2\right\rceil \left\lfloor J/2\right\rfloor }{4},1\right),\label{F-1}
\end{equation}
where $\left\lceil \cdot\right\rceil $ and $\left\lfloor \cdot\right\rfloor $
denote the integer valued ceiling and floor functions respectively
\footnote{These functions may be defined as: $\left\lceil z\right\rceil =\min\left(i\in\mathbb{Z}:\; z\leq i\right)$
and $\left\lfloor z\right\rfloor =\max\left(j\in\mathbb{Z}:\; z\geq j\right)$.%
}. If $J$ is odd then $\left\lceil J/2\right\rceil \left\lfloor J/2\right\rfloor =\left(J^{2}-1\right)/4$,
and $\left\lceil J/2\right\rceil \left\lfloor J/2\right\rfloor =J^{2}/4$
in the simpler case when $J$ is an even number. Note that the functions
(\ref{F-1}) form the increasing sequence as desired.

The major result of the above considerations is thus the family of
new majorization entropic uncertainty relations ($n=2,\ldots\infty$):

\begin{equation}
H_{\alpha}\bigl[q^{\Delta}\bigr]+H_{\alpha}\bigl[p^{\delta}\bigr]\geq\mathcal{R}_{\alpha}^{(n)}\left(\Delta\delta/\hbar\right)\equiv H_{\alpha}\bigl[W^{(n)}\left(\Delta\delta/\hbar\right)\bigr],\label{MEURCG}
\end{equation}
valid for $\alpha\leq1$. As mentioned in Section \ref{Secmajrev}
the case of the Shannon entropy directly follows from (\ref{add2}),
while the range $\alpha<1$ is obtained due to the subadditivity of
$\ln(1+z)$. In the case $\alpha>1$ we need to replace the majorization
bound according to (\ref{modified}), and obtain $\mathcal{R}_{\alpha}^{(n)}\left(\gamma\right)\equiv B_{\alpha\alpha}^{\alpha>1}\bigl[W^{(n)}\left(\gamma\right)\bigr]$.

\section{Discussion}

\begin{figure}
\includegraphics[scale=0.45]{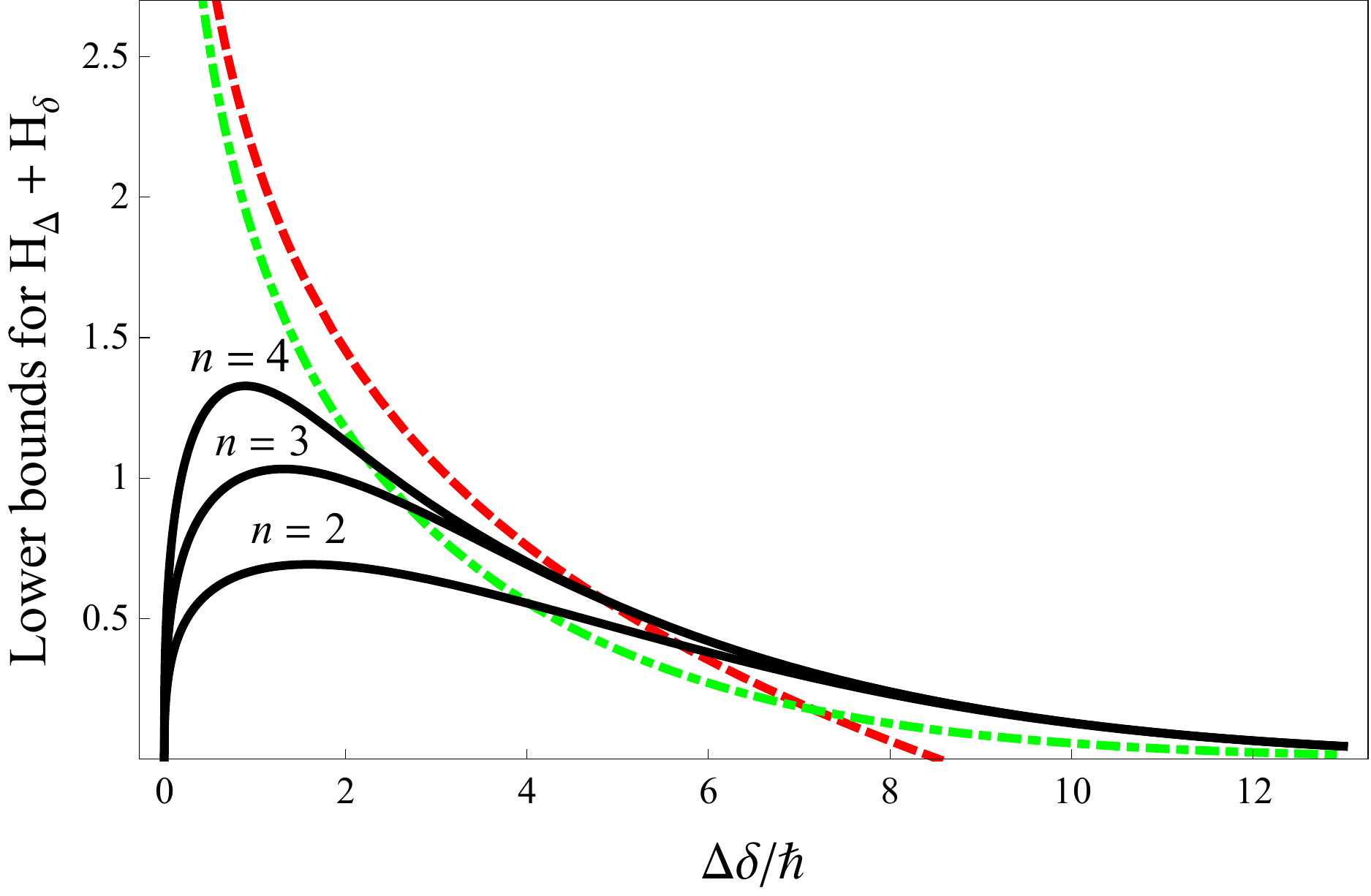}

\protect\caption{(Color online) As a comparison, I plot the previously known lower
bounds $\mathcal{B}_{1}$ (red dashed line), and $\mathcal{R}$ (green
dashed--dotted line), together with the new majorization bounds (black
solid lines), labeled by $n=2,3,4$. By $H_{\Delta}$ and $H_{\delta}$
I denote $H_{1}\bigl[q^{\Delta}\bigr]$ and $H_{1}\bigl[p^{\delta}\bigr]$
respectively. From the value $\Delta\delta/\hbar\approx4.8231$ in
which the black line ($n=4$) intersects the red dashed line, the
new bounds improve the previously known results.}
\end{figure}

The comparison of the previous bounds (\ref{IBB}) and (\ref{LR})
with the new majorization results is presented in Fig. 1, for the
case of the Shannon entropy ($\alpha=1=\beta$). I depicted the first
three majorization-based bounds (black, solid lines) since they are
sufficient to capture the whole content of the new uncertainty relations.
First of all, only the bound for $n=2$ is slightly weaker than the
remaining majorization bounds in the regime of larger $\gamma$, while
there is no difference between $n=3$, $n=4$ and other (not presented)
values of $n$. For $\gamma\rightarrow\infty$, all the black curves
exhibit the same behavior, so that one can take an advantage of the
asymptotic expansion \cite{Fuchs} 
\begin{equation}
\frac{\gamma}{2\pi}\left[R_{00}\left(\gamma/4,1\right)\right]^{2}\sim1-2\sqrt{\pi\gamma}e^{-\gamma/2},
\end{equation}
in order to show that 
\begin{equation}
\mathcal{R}_{1}^{(n)}\left(\gamma\right)\sim\frac{\sqrt{\pi}}{2}\gamma^{3/2}e^{-\gamma/2},\label{asy1}
\end{equation}
for all $n=2,\ldots\infty$. The same expansion studied for the previous
bound (\ref{LR}) leads to
\begin{equation}
\mathcal{R}\left(\gamma\right)\sim2\sqrt{\pi\gamma}e^{-\gamma/2}.\label{asy2}
\end{equation}
The asymptotic value of (\ref{asy1}) is larger than (\ref{asy2})
by a divergent factor $\gamma/4$. Since the bound (\ref{LR}) for
the sum of two Shannon entropies is always weaker than the couple
$\mathcal{B}_{1}\left(\gamma\right)$ and $\mathcal{R}_{1}^{(3)}\left(\gamma\right)$,
it is in this case sufficient to use only these two bounds. Obviously,
the bound $\mathcal{R}\left(\gamma\right)$ remains useful (as being
always non-negative) for the conjugated parameters $(\alpha,\beta)$
with $\alpha\neq\beta$, when the majorization bounds do not apply.
Let me remind, that in the limiting case $\alpha=1/2$, $\beta=\infty$,
the bound $\mathcal{R}\left(\gamma\right)$ is optimal and can be
saturated for any value of $\gamma$. 

While increasing the number $n$ we do not change the tail of the
bound, we still substantially improve the area of small $\gamma$.
Taking the limit $\gamma\rightarrow0$ one can recognize that the
optimal majorization bound $\mathcal{R}_{1}^{(\infty)}\left(\gamma\right)$,
behaves like $-\frac{1}{2}\ln\gamma$, so is still far below the bound
$\mathcal{B}_{1}\left(\gamma\right)$. To show that property one needs
to associate $i\sqrt{\gamma}$ in (\ref{Wi}) with a continuous variable
$z$, so that
\begin{equation}
W_{i}\left(\gamma\right)\rightarrow\sqrt{\gamma}\frac{d}{dz}\left[\frac{z}{2\sqrt{2\pi}}R_{00}\left(z^{2}/16,1\right)\right],\label{Wi-1}
\end{equation}
and use the definition of the Riemann integral. This kind of behavior
is somehow typical in the majorization approach to entropic uncertainty
relations. In the discrete case, the tensor-product EUR (weaker than
the direct-sum EUR used in this paper) can outperform the Maassen-Uffink
result in more than $98\%$ of cases \cite{my}, even for a small
dimension of the Hilbert space equal to $5$. But the Maassen-Uffink
lower bound \cite{MU} always dominates when $U_{ij}$ is sufficiently
close to the Fourier matrix, so that both eigenbases of the observables
$A$ and $B$ become mutually unbiased. The continuous limit $\gamma\rightarrow0$
is of exactly the same sort, since the resulting continuous densities
originate from the wave functions in position and momentum spaces,
which are related by the Fourier transformation. Note that the behavior
in the limit $\gamma\rightarrow0$ does not thus permit us to derive
counterparts of the continuous EURs \cite{BBM,IBB2,Vignat}, valid
for $\beta=\alpha$.

\section{Conclusions}

The direct-sum majorization entropic uncertainty relation for coarse-grained
observables given by Eq. (\ref{MEURCG}) is the first known bound
in the case $\beta=\alpha$ and $\alpha\neq1$. In the Shannon case,
the bound (\ref{Boundy}) holds as well and the comparison of all
bounds is depicted in Fig. 1. The new bounds (black, solid lines)
significantly improve the previously known results in the regime of
$\gamma\geq4.8231$ (this threshold value is an intersection point
between the red dashed line and the black line labeled by $n=4$).
Such regime of relevance ($\gamma\geq4.8231$) is of practical importance.
In \cite{ConEnt2}, entanglement of a two-mode Gaussian state has
been experimentally confirmed with the coarse-graining widths $\Delta=17\Delta_{1}$
and $\delta=15\delta_{1}$, where $\Delta_{1}=0.0250\textrm{mm}$
and $\delta_{1}/\hbar=1.546\textrm{mm}^{-1}$. To construct the entanglement
criteria, one needs to put $\gamma=\Delta\delta/2\hbar$ inside the
underlying uncertainty relation (the factor of $1/2$ comes from different
normalization of global quadratures), so that the above numbers boil
down to the value $\gamma=4.9279$. Even though, we observe a tiny
overlap between the regime in which the new EUR outperforms the previous
results and the parameters from \cite{ConEnt2}, for slightly larger
coarse graining, say $\gamma=7$, the value of the bound increases
by $60\%$ because $\mathcal{R}_{1}^{(3)}\left(7\right)/\mathcal{B}_{1}\left(7\right)=1.609$.
This however suggests, that with the new bound at hand, one could
improve the performance of the entanglement criteria and possibly
detect entanglement beyond the cases reported in \cite{ConEnt2}.
The better detection ability might become important while dealing
with multipartite entanglement \cite{Saboia}, since due to the increasing
number of degrees of freedom, the coarse-grained measurements might
appear to be the one feasible experimental method \cite{personal}.

In the discrete scenario with almost mutually unbiased bases the Maassen--Uffink
bound always outperforms the majorization approach. It however still
can be improved with the help of the monotonicity property of the
relative entropy \cite{Coles}, or by combining the former approach
with the majorization techniques \cite{RPZ}. In the continuous case
this type of analysis is far more difficult, since we actually do
not have in our disposal the unitary matrix $U$ such that $a_{k}=\sum_{l}U_{kl}b_{l}$
and $a_{k},b_{l}$ are the probability amplitudes reproducing (\ref{rs}),
\begin{equation}
q_{k}^{\Delta}=|a_{k}|^{2},\qquad p_{l}^{\delta}=|b_{l}|^{2}.
\end{equation}
From the beginning we deal with the \emph{per se} probabilities $q_{k}^{\Delta}$
and $p_{l}^{\delta}$. This fundamental difference can be overcome
if one introduces an additional degree of freedom \cite{arxivMoj,RusLas,OptCon}
corresponding to the orthonormal bases on the intervals $\left[k_{a}^{-}\Delta,k_{a}^{+}\Delta\right]$
and $\left[l_{b}^{-}\Delta,l_{b}^{+}\Delta\right]$. Even though,
this approach brings the valid unitary matrix $U$, the remaining
optimization required by \cite{Coles} becomes a challenging task.

\acknowledgments

It is my pleasure to thank Iwo Bia\l ynicki-Birula and Karol \.{Z}yczkowski
for helpful comments. Financial support by grant number IP2011 046871
of the Polish Ministry of Science and Higher Education is gratefully
acknowledged. Research in Freiburg is supported by the Excellence
Initiative of the German Federal and State Governments (Grant ZUK
43), the Research Innovation Fund of the University of Freiburg, the
ARO under contracts W911NF-14-1-0098 and W911NF-14-1-0133 (Quantum
Characterization, Verification, and Validation), and the DFG (GR 4334/1-1).


\begin{thebibliography}{References}
\bibitem{BBM} I. Bia\l ynicki-Birula and J. Mycielski, Commun. Math.
Phys. \textbf{44}, 129 (1975).

\bibitem{Deutsch} D. Deutsch, Phys. Rev. Lett. \textbf{50}, 631 (1983). 

\bibitem{MU} H. Maassen and J. M. B. Uffink, Phys. Rev. Lett. \textbf{60},
1103 (1988).

\bibitem{Wehner} S. Wehner and A. Winter, New J. Phys. \textbf{12},
025009 (2010). 

\bibitem{IBBLR} I. Bia\l ynicki-Birula and \L . Rudnicki \emph{Entropic
uncertainty relations in quantum physics} in \emph{Statistical Complexity},
ed. K D Sen (Berlin: Springer) pp 1\textendash 34 (2011).

\bibitem{deVicente} J. I. de Vicente and J. Sanchez-Ruiz, Phys. Rev.
A \textbf{77}, 042110 (2008).

\bibitem{deVicenteComm} G. M. Bosyk, M. Portesi, A. Plastino, and
S. Zozor, Phys. Rev. A\textbf{ 84}, 056101 (2011).

\bibitem{Partovi}  M. H. Partovi, Phys. Rev. A \textbf{84}, 052117
(2011).

\bibitem{my} Z. Pucha\l a, \L . Rudnicki, and K. \.{Z}yczkowski,
J. Phys. A \textbf{46}, 272002 (2013).

\bibitem{oni} S. Friedland, V. Gheorghiu, and G. Gour, Phys. Rev.
Lett. \textbf{111}, 230401 (2013).

\bibitem{Coles} P. Coles and M. Piani, Phys. Rev. A \textbf{89},
022112 (2014).

\bibitem{RPZ} \L . Rudnicki, Z. Pucha\l a, and K. \.{Z}yczkowski,
Phys. Rev. A \textbf{89}, 052115 (2014).

\bibitem{Korzekwa1} K. Korzekwa, M. Lostaglio, D. Jennings, and T.
Rudolph, Phys. Rev. A \textbf{89}, 042122 (2014).

\bibitem{Bosyk1} S. Zozor, G. M. Bosyk, and M. Portesi, J. Phys.
A \textbf{46}, 465301 (2013). 

\bibitem{Bosyk2} G. M. Bosyk, S. Zozor, M. Portesi, T. M. Os\'an, and
P. W. Lamberti, Phys. Rev. A \textbf{90}, 052114 (2014).

\bibitem{Bosyk3} S. Zozor, G. M. Bosyk, and M. Portesi, J. Phys.
A \textbf{47}, 495302 (2014).

\bibitem{Kaniewski} J. Kaniewski, M. Tomamichel, and S. Wehner, Phys.
Rev. A \textbf{90}, 012332 (2014).

\bibitem{Banach} R. Adamczak, R. Lata\l a, Z. Pucha\l a, and K. \.{Z}yczkowski,
arXiv:1412.7065 (2015).

\bibitem{Lewenstein} O. G\"uhne and M. Lewenstein, Phys. Rev. A\textbf{
70}, 022316 (2004). 

\bibitem{Partovi2} M. H. Partovi, Phys. Rev. A \textbf{86}, 022309
(2012).

\bibitem{Rastegin1} A. E. Rastegin, arXiv:1407.7333 (2014). 

\bibitem{ConEnt1} S. P. Walborn, B. G. Taketani, A. Salles, F. Toscano,
and R. L. de Matos Filho, Phys. Rev. Lett. \textbf{103}, 160505 (2009).

\bibitem{ConEnt2} D. S. Tasca, \L . Rudnicki, R.M. Gomes, F. Toscano,
and S. P. Walborn, Phys. Rev. Lett. \textbf{110}, 210502 (2013).

\bibitem{crypto} I. B. Damgaard, S. Fehr, R. Renner, L. Salvail,
and C. Schaffner, in Advances in Cryptography\textemdash CRYPTO, LNCS
(Springer, New York, 2007), Vol. 4622, pp. 360\textendash 378.

\bibitem{crypto2} N. H. Y. Ng, M. Berta, and S. Wehner, Phys. Rev.
A \textbf{86}, 042315 (2012).

\bibitem{Berta}  M. Berta, M. Christandl, R. Colbeck, J. M. Renes,
and R. Renner, Nat. Phys. \textbf{6}, 659 (2010).

\bibitem{memory2} T. Pramanik, P. Chowdhury, and A. S. Majumdar,
Phys. Rev. Lett. \textbf{110}, 020402 (2013).

\bibitem{Steering1}  J. Schneeloch, P. B. Dixon, G. A. Howland, C.
J. Broadbent, and J. C. Howell, Phys. Rev. Lett. \textbf{110}, 130407
(2013).

\bibitem{EPR} J. Schneeloch, C. J. Broadbent, and J. C. Howell, Phys.
Lett. A\textbf{ 378}, 766 (2014).

\bibitem{HeisWerner} P. Busch, P. Lahti, and R. F. Werner, Phys.
Rev. Lett. \textbf{111}, 160405 (2013).

\bibitem{HeisEntropic} F. Buscemi, M. J. W. Hall, M. Ozawa, and M.
M. Wilde, Phys. Rev. Lett. \textbf{112}, 050401 (2014).

\bibitem{Rastegin2}  A. E. Rastegin, arXiv:1406.0054 (2014). 

\bibitem{ColFur} P. J. Coles, F. Furrer, Phys. Lett. A \textbf{379},
105\textendash 112 (2015).

\bibitem{partoviOld} M. H. Partovi, Phys. Rev. Lett. \textbf{50},
1883 (1983).

\bibitem{IBB1} I. Bialynicki-Birula, Phys. Lett. A \textbf{103},
253 (1984).

\bibitem{IBB2} I. Bialynicki-Birula, Phys. Rev. A \textbf{74}, 052101
(2006).

\bibitem{Lenard} A. Lenard, J. Functional Analysis \textbf{10}, 410
(1972).

\bibitem{Lahti} P. Lahti, Rep. Math. Phys. \textbf{23}, 289 (1986).

\bibitem{Lahti2} P. Busch, T. Heinonen, P. Lahti, Physics Reports
\textbf{452}, 155 (2007).

\bibitem{OptCon} \L . Rudnicki, S. P. Walborn, and F. Toscano, Phys.
Rev. A \textbf{85}, 042115 (2012).

\bibitem{Megan} M. R. Ray and S. J. van Enk, Phys. Rev. A \textbf{88},
042326 (2013).

\bibitem{abr} M. Abramowitz and I. A. Stegun, \textit{Handbook of
Mathematical Functions}. Dover, New York, (1964).

\bibitem{LandauPollak} H. J. Landau and H. O. Pollak, Bell System
Tech. J. \textbf{40}, 65 (1961).

\bibitem{Price} M. Cowling and J. Price, SIAM J. Math. Anal. \textbf{15},
151 (1984).

\bibitem{Fuchs} W. H. J. Fuchs, J. Math. Anal. Appl.\textbf{ 9},
317 (1964).

\bibitem{Vignat} S. Zozor and C. Vignat, Physica A \textbf{375} 499
(2007).

\bibitem{Saboia} A. Saboia, A. T. Avelar, S. P. Walborn, and F. Toscano,
arXiv:1407.7248 (2014).

\bibitem{personal} D. Tasca and S. P. Walborn, personal communication,
(2014).

\bibitem{arxivMoj}\textbf{ }\L . Rudnicki, \emph{Uncertainty related
to position and momentum localization of a quantum state} in: ``\emph{Proceedings
of New Perspectives in Quantum Statistics and Correlations}'', M.
Hiller, F. de Melo, P. Pickl, T. Wellens, S. Wimberger (Eds.), Universitatsverlag
Winter, p. 49 (2012). 

\bibitem{RusLas} \L . Rudnicki, J. Russ. Laser Res. \textbf{32},
393 (2011).\end{thebibliography}
\end{document}